\documentclass[letterpaper]{article} 
\usepackage{aaai23}  
\usepackage{times}  
\usepackage{helvet}  
\usepackage{courier}  
\usepackage[hyphens]{url}  
\usepackage{graphicx} 
\usepackage{natbib}  
\usepackage{caption} 
\usepackage{algorithm}
\usepackage{algorithmic}

\usepackage{newfloat}
\usepackage{listings}
\DeclareCaptionStyle{ruled}{labelfont=normalfont,labelsep=colon,strut=off} 
\floatstyle{ruled}
\newfloat{listing}{tb}{lst}{}
\floatname{listing}{Listing}
\newcommand\rev[1]{#1}
\newcommand\qt[1]{``\textit{#1}''}
\usepackage{caption}
\usepackage{subcaption}

\usepackage[textsize=scriptsize,disable]{todonotes}

\newcommand{\acite}[1]{{\protect\citeauthor{#1}\protect}~(\citeyear{#1})}
\newcommand{\eg}{\textit{e.g.,}}
\newcommand{\ie}{\textit{i.e.,}}
\newcommand{\mturk}{MTurk}

\usepackage{colortbl}
\newcommand{\cc}{\cellcolor{blue!15}}
\newcommand{\cca}{\cellcolor{orange!45}}
\newcommand{\ccb}{\cellcolor{orange!30}}
\newcommand{\ccc}{\cellcolor{orange!15}}

\title{How Crowd Worker Factors Influence \\Subjective Annotations: A Study of Tagging Misogynistic Hate Speech in Tweets}
\author{
    Danula Hettiachchi,\textsuperscript{\rm 1}
    Indigo Holcombe-James,\textsuperscript{\rm 1}
    Stephanie Livingstone,\textsuperscript{\rm 1}
    Anjalee de Silva,\textsuperscript{\rm 2}
    Matthew Lease,\textsuperscript{\rm 3}
    Flora D. Salim,\textsuperscript{\rm 4}
    Mark Sanderson \textsuperscript{\rm 1}
}
\affiliations{
    \textsuperscript{\rm 1} RMIT University, Australia, 
    \textsuperscript{\rm 2} The University of Melbourne, Australia, \\
    \textsuperscript{\rm 3} University of Texas at Austin, United States, 
    \textsuperscript{\rm 4} University of New South Wales, Australia\\
    danula.hettiachchi@rmit.edu.au, indigo.holcombe-james@rmit.edu.au, stephanie.livingstone@student.rmit.edu.au, adesilva@unimelb.edu.au, ml@utexas.edu, flora.salim@unsw.edu.au, mark.sanderson@rmit.edu.au
}

\begin{document}

\maketitle

\begin{abstract}
Crowdsourced annotation is vital to both collecting labelled data to train and test automated content moderation systems and to support human-in-the-loop review of system decisions. However, annotation tasks such as judging hate speech are subjective and thus highly sensitive to biases stemming from annotator beliefs, characteristics and demographics. We conduct two crowdsourcing studies on Mechanical Turk to examine annotator bias in labelling sexist and misogynistic hate speech. Results from 109 annotators show that annotator political inclination, moral integrity, personality traits, and sexist attitudes significantly impact annotation accuracy and the tendency to tag content as hate speech. In addition, semi-structured interviews with nine crowd workers provide further insights regarding the influence of subjectivity on annotations. In exploring how workers interpret a task --- shaped by complex negotiations between platform structures, task instructions, subjective motivations, and external contextual factors --- we see annotations not only impacted by worker factors but also simultaneously shaped by the structures under which they labour.
\end{abstract}

\section{Introduction}

Social media platforms allow users to create and share content with one another at an unprecedented scale. 
While this provides an incredibly powerful channel for human communication, the lack of accountability and the possibility to remain anonymous on online platforms have resulted in various online safety concerns, such as toxic language in various forms \cite{Fortuna2018-qe,MacAvaney2019-gw,Poletto2021-fl,Schmidt2017-oa,Vidgen2020-fu} (\eg~hate speech, abusive language, offensive language, etc.). Robust and reliable automated tooling, coupled with human-in-the-loop review, play a vital role in moderating online content to help create safe online spaces. Prediction models typically perform the initial review, with automatic decisions in cases of high-confidence, and less confident cases deferred to human review. As the adage goes, ``garbage in equals garbage out''~\cite{Vidgen2020-fu}: to train reliable data-driven machine learning models, we must first obtain high-quality training and testing data~\cite{Aroyo2022-hr,Sambasivan2021-ps}.

However, annotation tasks like judging hate speech and classifying misinformation often require subjective judgement, which is potentially impacted by biases held by annotators~\cite{Aroyo2019-om}. For example, \acite{Hube2019-nv} show how worker factors influence outcomes in a task that asks participants to label statements on controversial topics as `neutral' or `opinionated'. Similarly, a broader range of factors, such as skills~\cite{Kumai2018-bv}, cognitive ability~\cite{Hettiachchi2020-bc}, and worker environment~\cite{Gadiraju2017-le} influence task outcomes in both objective and subjective crowdsourcing tasks. In addition to worker factors, clarity of the task~\cite{Gadiraju2017-fy}, user interface~\cite{Alagarai_Sampath2014-eh}, and incentives~\cite{Singer2013-he} could also create biases. Whether originating due to worker factors, or inflicted by requester actions, these biases can significantly impact the annotation data quality with broader implications for models trained using labelled data~\cite{Bender2021-en,Wiegand2019-nr,Smith2022-fn}. 

Significant prior work has examined agreement among crowd workers, comparing them with expert annotations, and exploring the relationships between worker characteristics and annotation quality~\cite{Poletto2021-fl,Wiegand2019-nr}. For example, recent work on toxic language annotation by ~\acite{Sap2022-qg} investigates how annotator beliefs regarding race influence their toxicity ratings of African American English text. However, examining the relationship between questionnaire and annotation outcomes provides only limited insights into the worker's sense-making process, as well as how and whether they attempt to alienate personal beliefs to achieve consistency with task instructions and expectations. 

To address this, we conducted two crowdsourcing studies on Amazon Mechanical Turk (\mturk) that combine quantitative and qualitative methods to better investigate how annotator biases may influence tagging of sexist and misogynistic hate speech on Twitter. While much prior work has explored worker biases in subjective judgements relating to hate speech tagging at large, our work specifically investigates misogynistic or sexist hate speech on the Twitter platform. Our first study gathers quantitative input from workers performing hate speech annotation. Results from 109 annotators show that annotator political inclination, moral integrity, personality traits, and sexist attitudes significantly impact annotation accuracy measured through agreement with expert annotations. Similarly, worker attributes have significant impacts on what portion of tweets they tag as hate speech. In our second study, we conducted semi-structured interviews with 9 crowd workers to garner further insights into how work practices and context, including employment conditions, can contribute to bias in subjective annotation.

{\bf Contributions.} We make three key contributions. First, our study reveals a significant impact of political inclination, moral integrity, personality traits, and sexist attitudes of workers on their accuracy in annotating misogynistic hate speech. Second, through drawing on qualitative, semi-structured interviews with workers as they undertook their annotations, our findings offer new insights into the complex negotiations of the platform and personal conditions, structures and standards that annotation work requires. By highlighting how crowd work and the resulting annotations are influenced not only by worker factors but the structures under which they labour, we evidence the utility of a more holistic accounting of influential factors, and open the potential for more nuanced design considerations in the future. Finally, in collaboration with subject matter experts, we create and share a new Twitter dataset that contains misogynistic or sexist hate speech, including expert annotations of five subcategories of hate speech.

\section{Related Work}

\subsection{Subjective Annotation Tasks}
Annotation tasks span a wide extreme from {\em objective} tasks (\eg~is there a dog in this picture?) to {\em subjective} tasks (\eg~what is your favourite colour?), with a wide, varying-range of partially-subjective tasks \cite{Nguyen2016-od} in between. Collecting annotations for subjective phenomena, in general, is challenging, and more so in crowdsourcing contexts involving temporary, contingent work between unknown parties with minimal communication channels. 

\acite{Aroyo2019-om} discuss two types of subjectivity that can impact the worker judgement in toxicity annotation. First, subjectivity could be inherent to the topic or the domain, where workers could draw on personal preference or experience to make the judgement. In these tasks, two workers could simply have different opinions. Second, subjectivity could occur due to the ambiguity in input items or task instructions. 
Similarly,~\acite{Sen2015-sk} examine semantic relatedness judgements that ask workers to rate the strength of the relationship between two concepts. The study on \mturk\ reports that annotated datasets collected from 39 \mturk\ crowd annotators significantly vary from annotations collected from 72 scholars. Similar results were evident from work by~\acite{Hube2019-nv}, who found that crowd workers with strong opinions produce biased annotations, even evident among experienced workers. Furthermore, demographics, location, worker context, and work environment have been found to impact annotation quality in both subjective and objective tasks~\cite{Hettiachchi2021-gf}.

\subsection{Mitigating Bias in Crowdsourced Annotation} 

Various approaches exist for detecting and mitigating biases stemming from crowdsourced annotations. While numerous post-annotation approaches exist to detect and/or mitigate algorithmic bias, it is preferable to avoid or reduce biases during the annotation process itself. Common data quality control measures, such as comparing worker labels to gold standard labels (\eg~from experts), have typically been studied in the context of objective labelling tasks in which disagreement with gold can be construed as a labelling error. While the labelling error in objective tasks could represent the overall annotation quality, with subjective tasks, limited error measures cannot accurately cover the broader spectrum of subjective judgements involved in a task.
Researchers have thus explored alternative methods, such as task design and presentation strategies. \acite{Hube2019-nv} experimented with a task where workers are given statements related to controversial topics and asked to tag them as `neutral' or `opinionated'. Their results show that social projection -- or asking workers to label according to how they believe the majority of other workers would label them -- leads to more consistent annotations. Similarly, ``awareness reminders'' ask workers to reflect on the controversial nature of the topics in the task and be mindful of the potential bias their personal opinions could have on their judgements. 

Other approaches to mitigate bias require task requesters (\eg~researchers or machine learning practitioners) to manually specify population requirements to mitigate bias. For example, \acite{Barbosa2019-ln} allow requesters to specify whether they need balanced or skewed populations with respect to specific worker characteristics. While such selections can help task requesters avoid certain worker groups depending on task needs, determining appropriate population parameters is not always straightforward. Task requesters themselves may also embody their own biases, complicating matters even further. 

\subsection{Implications of Precarious Work}

Online crowd work, or data work more generally, has well-known challenges, leading to characterisation as {\em invisible work}~\cite{Irani2013-qz} or {\em ghost work}~\cite{Gray2019-sb}. While researchers and practitioners have sought a ``gold mine'' of inexpensive labelling of gold data, others have suggested that a ``coal mine'' metaphor may be more appropriate \cite{Fort2011-uc}. 
The negative impacts of these forms of labour on workers and their ethical consequences have a long history of such critique \cite{Vallas2020-my, Felstiner2011-ao}. The most salient consequence of this, in regard to this particular study, is understanding the extent to which such precarious work can influence the resulting work products by undermining the quality and/or creating or amplifying biases in annotation. For example, the question of who is willing to undertake such work in general, or at the wages offered, is known to shape the demographics of the annotators who choose to participate in research studies or data labelling work. 

The piecemeal nature of the work and its low pay rates ensure earning a living wage is challenging~\cite{Berg2021-wl,Naderi2018-wq,Alkhatib2017-cy}. Further, through breaking tasks down, and dispersing their completion, crowd work has enabled some firms to move away from reliance on in-house employment~\cite{Berg2015-by}, exacerbating the ethical considerations surrounding such modes of employment~\cite{Vallas2020-my}. These considerations are compounded by the power dynamics that frame this work, with task requesters able to reject and refuse payment for any tasks not completed to their satisfaction~\cite{Rea2020-wc,Alkhatib2017-cy}. Qualitative research has been critical in highlighting these dynamics, particularly in drawing attention to how platforms such as \mturk\ ``rely not only on the calculative mechanisms of control that metrics afford but also on normative mechanisms in the form of [...] inducements that strengthen user attachment''~\cite{Vallas2020-my}(p. 279). Examining how these ``hierarchical structures [...] inform the interpretation of data''~\cite{Miceli2020-qt} (p. 2) is therefore critical.

\subsection{Hate Speech and its Annotation}

Hate speech annotation represents a family of subjective annotation tasks~\cite{Fortuna2018-qe} essential to both offline data collection for model training and testing, as well as online, human-in-the-loop content moderation.  
Despite online platforms (\eg~\cite{Meta2023-cx, Twitter2023-ce}) and authorities having recognised hate speech as a growing concern, the definition of hate speech lacks global consensus, while many different definitions are used for related and equivalent concepts~\cite{Fortuna2018-qe,Poletto2021-fl}. This lack of consensus, and the resulting variance in data annotation practices, also makes it more challenging to conduct rigorous benchmarking evaluations across publicly available datasets~\cite{Fortuna2020-el}.

As with other annotation tasks, bias in training data could lead to biased and inaccurate models~\cite{Bender2021-en,Wiegand2019-nr,Smith2022-fn}. \acite{Wiegand2019-nr} highlight that under more realistic settings, abusive speech classification performance is much lower than what is reported in previous research.
Similarly, recent work shows that agreement among annotators strongly correlates with hate speech recognition quality in automated methods~\cite{Kocon2021-og}. \acite{Rahman2021-pt} report a significant drop in hate speech detection accuracy of current models when tested on a new hate speech dataset that includes broader forms of hate speech. \acite{Sap2022-qg} examine how the annotator beliefs and identities inflict biases in toxic language detection tasks. In particular, \citeauthor{Sap2022-qg} report notable variations in judgements considering conservative annotators and ones scoring higher values for racist beliefs. In their task, such annotators were less likely to rate anti-Black language as toxic, but more likely to rate African American English as toxic. \rev{\acite{Salminen2018-ga} report on a study where crowd workers of 50 countries provided toxicity annotations for social media comments. The reported hate interpretation (\ie~rated intensity of hatefulness) scores differ significantly between the countries considered, but the interpretation differs more by individuals than by countries. \acite{Salminen2018-ga} note the importance of considering user attributes in the hate speech annotation processes and studying their impact.}

A wide variety of hate speech datasets now exist~\cite{Poletto2021-fl,Vidgen2020-fu, Davidson2017-kq}. 
\acite{Poletto2021-fl} review 64 datasets (56 annotated corpora and 8 lexica), with only 9 including sexist or misogynistic content. In terms of sources, 32 datasets contain tweets, while others include comments from social platforms like Facebook, Reddit, and YouTube, as well as news websites. The annotation process involves experts (\ie~judges with subject knowledge), or non-expert volunteers or annotators recruited through crowdsourcing platforms. 
In general, broad coverage hate speech datasets typically contain a relatively small amount of misogynistic or sexist hate speech.

\acite{Guest2021-pn} propose a taxonomy for classifying misogynistic hate speech and a dataset based on Reddit posts and comments. The paper defines four categories for misogynistic content: {\em Misogynistic Pejoratives}, {\em Descriptions of Misogynistic Treatment}, {\em Acts of Misogynistic Derogation}, and {\em Gendered Personal Attacks Against Women}. Categories for non-misogynistic content include {\em Counter-Speech Against Misogyny}, {\em Non-misogynistic Personal 
Attacks}, and other content that doesn't belong to any of the categories. \rev{While many other hate speech taxonomies have been proposed in the literature~\cite{Fortuna2018-qe}, they provide limited coverage on specific sub-categories needed to distinguish different types of misogynistic hate speech~\cite{Guest2021-pn}.}

Increasing attention is being devoted to how annotator attributes influence the annotation decisions in assessing hate speech at large  \cite{Al_Kuwatly2020-pm,Keswani2021-li,Kumar2021-bu,Sap2022-qg,Gordon2022-zw}. However, relatively little work has explored the impact of annotator attributes on sexist or misogynistic hate speech annotation decisions in particular \cite{Wojatzki2018-lk,Goyal2022-hk}. \acite{Wojatzki2018-lk} report a study with 40 men and 40 women judging 400 German assertions about women, and shows that both men and women judge hate speech consistently, particularly for cases of extreme misogyny. Civil Comments \cite{Borkan2019-sr} provides toxicity labels for 1.8M news comments, with 450K comments also labeled for the demographic group targeted (\eg~gender). In forthcoming work, \acite{Goyal2022-hk} augment the Civil Comments dataset with new annotations for annotator demographics, thereby allowing the study of how annotator demographics impact labelling decisions for different target groups.

\section{Study}

\begin{figure*}[htb]
    \centering
    \includegraphics[width=.8\textwidth]{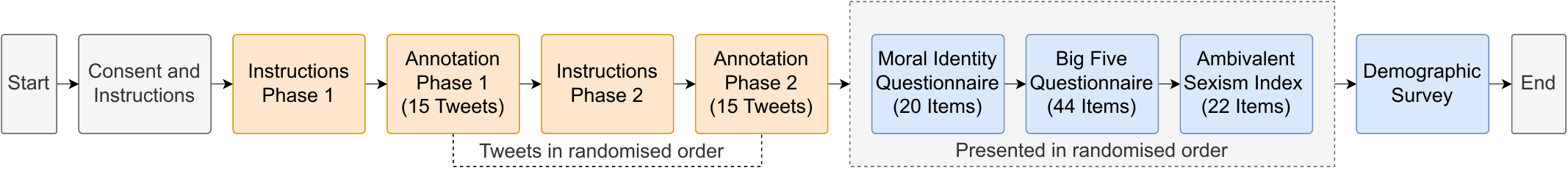}
    \caption{Flow of Study 1.}
    \label{fig:study-flow}
\end{figure*}

To better understand how annotator biases may influence annotation of sexist and misogynistic hate speech on Twitter, we conduct two crowdsourcing studies on \mturk\ that combine quantitative and qualitative methods. First, we gather sexist and misogynistic hate speech labels from crowd annotators in order to measure effects quantitatively. Following this, we conduct semi-structured interviews with nine crowd workers to garner further insights into their work context and thought processes.  
We first discuss common elements across the two studies, then present the details of each study in the following sections.
Both studies were approved by our human subjects oversight board.

\subsection{Sampling Tweets for Annotation}
\label{sec:dataset}

Hate speech is a broad term that refers to content targeting a person or group of people on the basis of their protected characteristics (\eg~race, ethnicity, national origin, disability, religious affiliation, sexual orientation, sex, gender identity). In this work, we consider annotation of sexist/misogynistic hate speech or hate speech directed at women on Twitter, a specific target group which has received far less attention than the study of hate speech annotation at large. 

A subject expert on anti-discrimination, and free speech and media law curated a list of 16 Twitter profiles (public figures) susceptible to sexist and misogynistic hate speech. We also contacted these individuals to obtain their consent to include their Twitter handle and relevant tweets in our research. We received positive responses from seven individuals, and collected the 5,000 most recent tweets for each profile. The selected individuals included authors, journalists, and activists from White, African-American, and Asian racial groups located in the US, UK, and Australia. Their profiles had 194,308 followers and 26,098 tweets on average. To avoid ambiguity on whether a tweet is directed at the specific person, we excluded any tweet that included more than one mentions (\ie~twitter handles in the tweet). Finally, we ordered by the tweet date and selected 2000 recent tweets for each profile. 

\subsection{Annotation Guidelines and Expert Labels}
\label{sec:guidelines}

Prior work has examined the utility of categories/factors, rating scales and comparative judgements when collecting hate speech annotations~\cite{Poletto2019-mo}. While rating scales are reported to be easy to use for annotators, they produce results which are difficult to understand and lead to poor overall performance in automated models. Comparative judgements such as best-worst-scale lead to better model outcomes. However, such ranking input scales are not suitable for our study design. Therefore, we collect categorical responses in our study. \rev{As we are interested in examining a specific type of hate speech, we use a tailored hate speech taxonomy instead of using generic and broader hate speech taxonomies~\cite{Salminen2018-kk} that have limited and incomplete label categories for misogynistic hate speech.} Specifically, \acite{De_Silva2020-nl} presents a functional theory of sex-based vilification, and \acite{Guest2021-pn} propose a taxonomy, from which we adapt the following categories of hate speech:
\begin{itemize}
    \item \textit{Threats and Violent Abuse (TVA)} are particularly serious examples of sex-based hate speech. It includes death and rape threats, as well as violent `correctives' which tell the participant that they `deserve/need [to be violated]' 
    \item \textit{Sexualised Abuse (SA)} reduces women to tools or objects for men's sexual use or pleasure. It exposes women's sexuality in public, irrelevant, humiliating, and/or distressing ways.
    \item \textit{Other Objectifying Speech (OOS)} is speech that is less severe than threats and violent abuse or sexualised abuse, but that still treats women as for use by others (typically by men). It is speech that treats women as valuable only if they are attractive or useful to others. 
    \item \textit{Other Contemptuous Speech (OCS)} communicates hatred, dislike, or disrespect for women that does not fit the categories described above. Other contemptuous speech typically treats women as inferior to men.
    \item \textit{Other Hate Speech (OHS)} is content that you think ought to be tagged as hate speech, but that does not fit into any of the above categories of sex-based hate speech.
\end{itemize}

Two authors (one woman, and one man) independently annotated \rev{the selected collection containing 14,000 tweets} according to the annotator instructions and examples generated for the annotation Phase 2 (\ie~five hate speech categories and not hate speech). Afterwards, a third author, the expert mentioned earlier, reviewed the annotations, discussed any conflicts with the other two authors, and determined the final gold annotations. For the final dataset, we selected all the tweets tagged as hateful, and randomly selected a subset of tweets tagged as not hateful. Our final dataset includes 140 tweets, with 90 tagged as not hateful and 50 tweets tagged as hate speech (Threats and violent abuse - 4, Sexualised abuse - 9, Other objectifying speech - 7, Other contemptuous speech - 10, Other Hate Speech - 20). \rev{While we treat these expert annotations as gold data to evaluate the agreement between annotators and experts, we acknowledge that expert annotations are not equivalent to ground truth due nuanced nature of defining hate speech~\cite{Fortuna2020-el}}.

\subsection{Study 1: Collecting Hate Speech Annotations}
\label{sec:study-1}
\label{sec:interface}

The first \mturk\ study compared hate speech labelling decisions of crowd workers under two variant designs: 1) binary labelling (`Hate Speech' or `Not Hate Speech') vs.\ 2) fine-grained labelling between the five specific types of hate speech described in Section \ref{sec:guidelines} vs.\ `Not Hate Speech'. 
As illustrated in {\bf Figure \ref{fig:study-flow}}, each worker first performed binary labelling of 15 tweets (Phase 1), and then performed fine-grained labelling of another, different set of 15 tweets (Phase 2).
For each available category, task instructions both described the category and provided two example tweets. Workers were shown the instructions before beginning each phase. Instructions could be reviewed at any time during annotation by clicking the `Show Instructions' button.

Following annotation, workers were presented with three questionnaires (in randomised order) and a final demographic survey, as shown in Figure~\ref{fig:study-flow}. Similar to \acite{Hube2019-nv}'s ``awareness reminders'', some prior work has shown that asking participants about their demographics at the start of a study can increase their self-awareness and influence their later responses, which is why we left this demographic survey for the very end. Self-reported demographic attributes collected include gender, political identity, the average weekly income, and crowdsourcing income as a percentage of primary income. 

The study included three questionnaires. The 44-item {\em Big Five Personality} questionnaire~\cite{Goldberg1992-fm} has been used in prior research that investigates annotation quality in crowd work~\cite{Lykourentzou2016-jf,Kazai2012-le}. It measures five personality traits: Extroversion, Agreeableness, Conscientiousness, Neuroticism, and Openness. More pertinent to our specific hate speech annotation task, we also included the {\em Moral Identity} questionnaire ~\cite{Black2016-nw} and the {\em Ambivalent Sexism Inventory} questionnaire~\cite{Glick1996-dv}. The Moral Identity questionnaire primarily captures how people make moral choices and has two dimensions: moral identity and moral self. Moral identity measures ``the desire to make intention and action consistent, and how much value participants place on acting according to moral principles.''. Moral self measures ``how closely participants identify with moral values''. The Ambivalent Sexism Inventory measures hostile sexism and benevolent sexism~\cite{Glick1996-dv}. 

The task was deployed for \mturk\ using psiTurk (https://psiturk.org/). As pre-qualifications, workers were required to have completed more than 1000 tasks, have an approval rate greater than 95\%, be located in the US, and have \mturk's adult qualification. For quality control purposes, we also included two attention-check questions in the annotation task. This led to the exclusion of responses from 9 workers who failed at least one of the checks. 

\subsection{Study 2: Conducting Semi-structured Interviews}
\label{sec:study-2}

Our second \mturk\ study draws on previous approaches to ethnographic research with crowd workers such as that outlined by~\acite{Gray2016-fi} and ~\acite{Williams2019-wa}. We conducted semi-structured interviews via remote video meetings. Similar to Study 1, workers were required to have completed more than 1,000 tasks, have an approval rate greater than 95\%, be located in the US, and have \mturk's adult qualification. In addition, we did not allow any workers who had completed Study 1 to also attempt Study 2. Workers initially expressed interest and reserved a time slot through a qualification task and then started the interview task during the agreed time slot. 

Following ~\acite{Gray2016-fi}, we asked workers to ``demonstrate how they [do] their crowd work'' by sharing their screen with the researcher and talking through the decisions that underpin their work practices. They were asked to describe how and why they make certain decisions to better understand their thought processes, whether individually or structurally determined. 
Interviews lasted approximately 45 minutes, and participants were compensated \$15 USD.

\section{Results}

\subsection{Quantitative Results}

\rev{We use two standard measures in our analysis. First, we assess whether crowd worker annotations agree with our expert annotations, termed as ``accuracy'' when considering expert annotations as gold standard data. Second, we examine the ``percentage or the number of hate speech annotations'' provided by each annotator or received by a collection of tweets}.
We collected 8-12 annotations per tweet, with a total of 3,270 responses collected from 109 workers. On average, each worker spent 18.4 ($SD=10.4$) minutes on completing the full task. Workers spent 14.3 ($SD=5.2$) seconds per tweet on average in Phase 1. In Phase 2, the average annotation time was 15.2 ($SD=2.3$) seconds.

\subsubsection{Identifying Hate Speech}

\begin{figure}[t]
    \centering
    \includegraphics[width=.7\columnwidth]{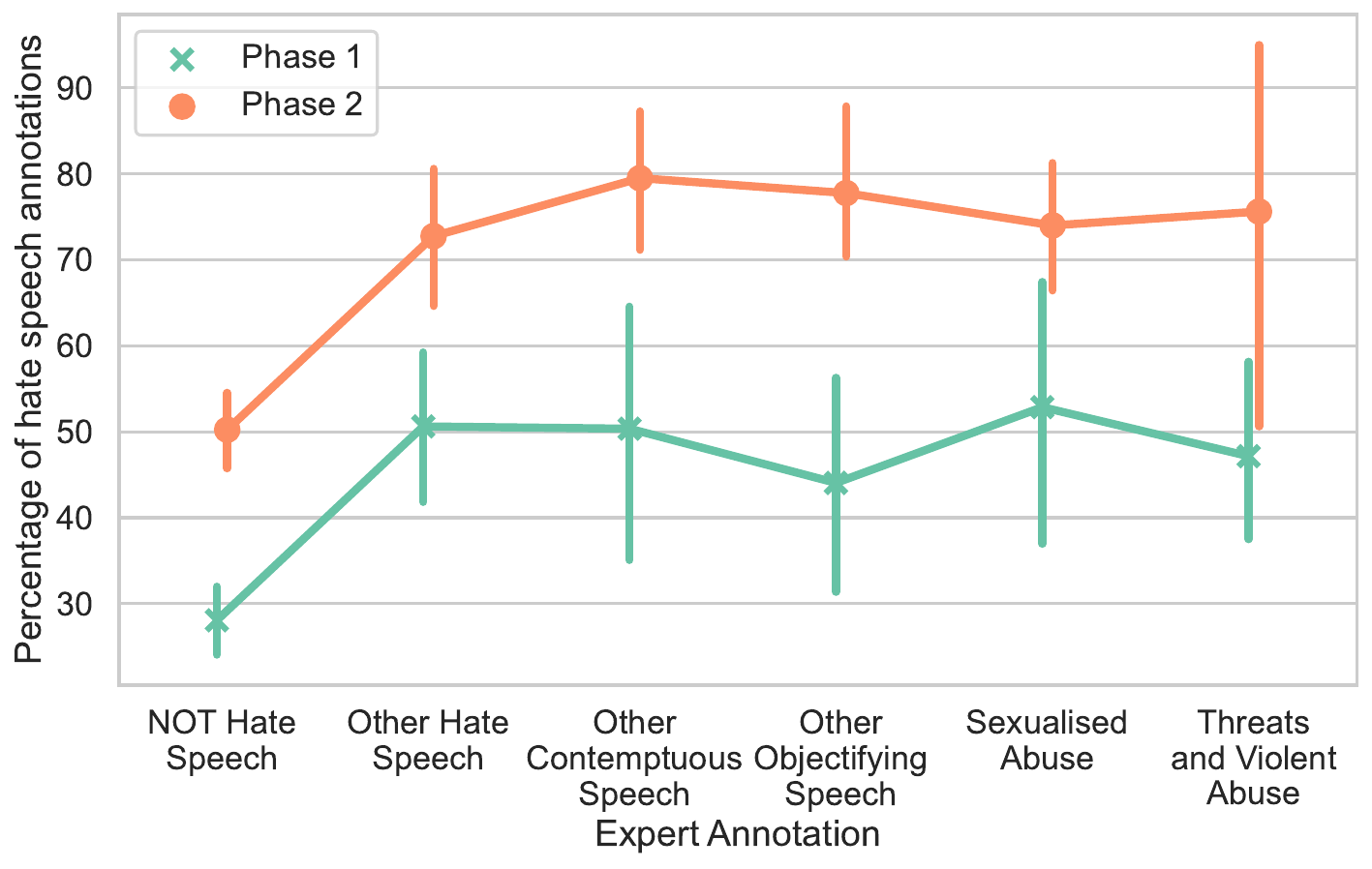}
    \caption{Percentage of hate speech annotations received for tweets categorised by the expert annotation.}
    \label{fig:tweets-category}
\end{figure}

How well do workers identify whether a tweet should be tagged as hate speech? For each tweet, we calculated the percentage of workers who rated it as hate speech (considering only binary hate / non-hate distinctions). {\bf Figure~\ref{fig:tweets-category}} shows the percentage of hate speech annotations received for each tweet group, categorised according to the expert annotation. 

\rev{In addition, to measure inter-annotator agreement, considering the nature of our crowdsourced data collection task, we use Krippendorf's alpha~\cite{Hayes2007-um}. Unlike other inter-annotator agreement measures, Krippendorf's alpha is suitable when each annotator only labels a subset of total items~\cite{Hayes2007-um}. When considering binary annotation outcomes, our results indicate a Krippendorf's alpha score of 0.149 in Phase 1 and 0.161 in Phase 2. While these scores only indicate slight agreement, it is a common observation in hate speech annotation tasks~\cite{Fortuna2018-qe}.}

\subsubsection{Categorising Hate Speech}

In the second phase of annotation, workers were expected to label the specific type of hate speech. Workers found it challenging to do so. As detailed in {\bf Table~\ref{tab:annotations-category}}, for example, for tweets with an expert annotation of `Threats and Violent Abuse (TVA)', only 9 (23.7\%) out of 38 annotations accurately labelled the specific category. Similarly, for `Other Contemptuous Speech (OCS)', only 33 (30.8\%) out of 107 annotations correctly classified the tweets. \rev{When considering all the categories in Phase 2, the resulting Krippendorf's alpha score of 0.086 also indicates very low agreement among workers.}

\begin{table}[t]
\footnotesize
\begin{center}
\begin{tabular}{lrrrrrrr}
\hline
\multicolumn{2}{c}{Crowd Worker}&\multicolumn{6}{c}{Expert Annotation}\\
\multicolumn{2}{c}{Annotation}& NHS & OHS & OCS & OOS & SA & TVA \\
\hline
Phase 1 & NHS &  \cc 702 & 105 & 54 & \cc 46 & 47 & \cc25\\
&HS & 268 & \cc112 & \cc55 & 36 & \cc54 & 22\\
\hline
Phase 2 & NHS & \cca496 & \cca62 & \ccb22 & \cca16 & \cca26 & \ccb8\\
&OHS & \ccc128 & \ccb52 & \ccc17 & 11 & 9 & \ccc7\\
&OCS & \ccb151 & \ccc42 & \cca33 & \ccc14 & \ccc16 & 5\\
&OOS & 103 & 34 & \ccc17 & \cca16 & \cca26 & 4 \\
&SA & 63 & 7 & 11 & 8 & \ccc16 & 5\\
&TVA & 50 & 21 & 7 & 6 & 8 & \cca9 \\
\hline
\end{tabular}
\caption{Total number of annotations received for tweets grouped by the expert annotation.} 
\label{tab:annotations-category}
\end{center}
\end{table}

\subsubsection{Relationship with Questionnaire Outcomes}

We calculated annotation outcomes for each worker to examine the impact of worker factors. Shapiro-Wilk tests indicated that accuracy and hate speech annotation percentage scores do not follow a normal distribution. Spearman rank correlation coefficients between questionnaire outcome scores, and worker accuracy and percentage of hate speech annotations provided by each worker are reported in {\bf Table~\ref{tab:correlations}}. We observe strong negative correlations between moral integrity and worker accuracy in Phase 2. For moral self, the results indicate a moderate positive correlation with Phase 2 accuracy. When examining relationships with the big five dimensions, three attributes (Agreeableness, Conscientiousness and Openness) have moderate positive correlations with Phase 2 accuracy. Furthermore, a moderate negative correlation is evident between benevolent sexism and worker accuracy in Phases 1 and 2.

\begin{table*}[htb]

\footnotesize
\begin{center}
\begin{tabular}{llrrrrrr}
\hline
&& \multicolumn{3}{c}{with Accuracy} & \multicolumn{3}{c}{with \% of Hate Annotations} \\
Test & Dimension & Phase 1 & Phase 2 & Overall & Phase 1 & Phase 2 & Overall\\
\hline
Moral Identity  & Moral-Self & 0.234 & \textbf{0.384} & \textbf{0.384} & -0.154 & \textbf{-0.360} & \textbf{-0.312}\\
Questionnaire    & Moral-Integrity & \textbf{-0.480} & \textbf{-0.551} & \textbf{-0.607} & \textbf{0.368} & \textbf{0.614} & \textbf{0.567}\\
\hline
Big Five  & Extraversion & \textbf{-0.327} & -0.152 & -0.271 & 0.286 & 0.183 & 0.226 \\
Personality& Agreeableness & \textbf{0.302} & \textbf{0.443} & \textbf{0.459} & -0.257 & \textbf{-0.470} & \textbf{-0.435}\\
& Conscientiousness & 0.175 & \textbf{0.381} & \textbf{0.350} & -0.220 & \textbf{-0.484} & \textbf{-0.428}\\
& Neuroticism & -0.083 & -0.154 & -0.129 & 0.123 & 0.283 & 0.250\\
& Openness & 0.111 & \textbf{0.338} & 0.298 & -0.158 & \textbf{-0.326} & -0.286\\
\hline
Ambivalent   & Hostile & -0.271 & -0.057 & -0.173 & 0.117 & 0.125 & 0.132 \\
Sexism Inventory& Benevolent & \textbf{-0.446} & \textbf{-0.364} & \textbf{-0.467} & \textbf{0.337} & \textbf{0.496} & \textbf{0.477} \\
\hline
\end{tabular}
\end{center}
\caption{Spearman rank correlation scores between questionnaire outcomes and hate speech annotations.}
\label{tab:correlations}
\end{table*}

\rev{When considering the percentage of hate speech annotations, in most cases, the correlation scores are in the opposite direction compared to accuracy. Between moral integrity and the percentage, strong positive correlations are evident in Phase 2, while a moderate positive correlation exists in Phase 1. We note a moderate negative correlation between moral self and the percentage in Phase 2. With the big five dimensions, Agreeableness and Conscientiousness have moderate negative correlations with Phase 2 accuracy. Moderate positive correlations are also evident between benevolent sexism and the percentage in both phases.}

\subsubsection{Impact of Demographic Factors}

\begin{figure}[!htb]
     \centering
     \begin{subfigure}[b]{0.48\columnwidth}
         \centering
         \includegraphics[width=.9\textwidth]{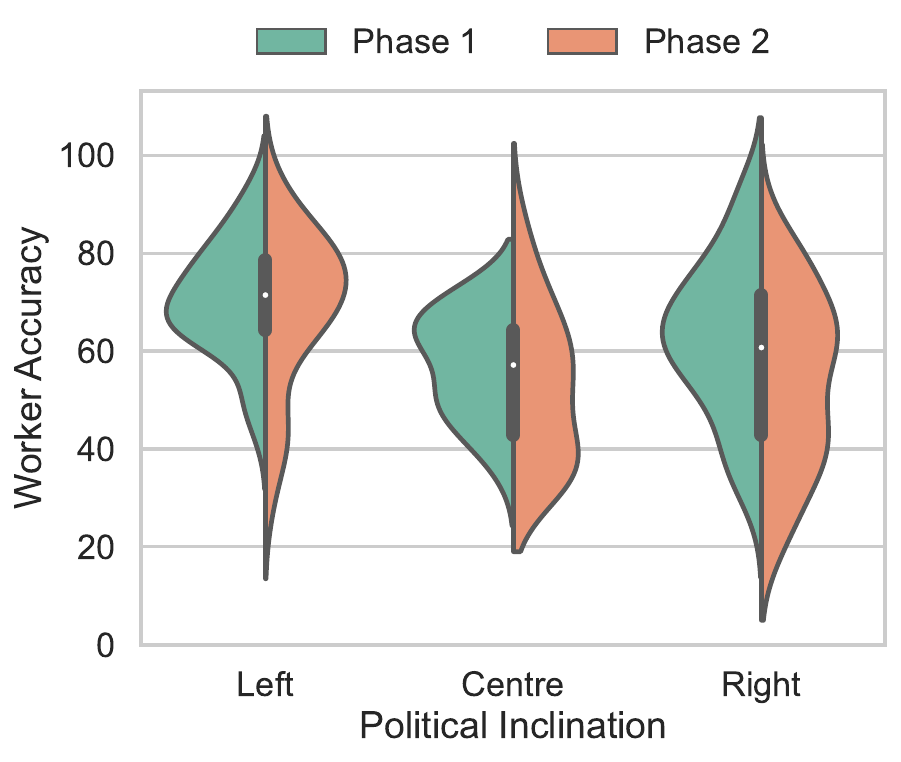}
         \caption{}
         \label{fig:political}
     \end{subfigure}
    \begin{subfigure}[b]{0.48\columnwidth}
         \centering
         \includegraphics[width=.9\textwidth]{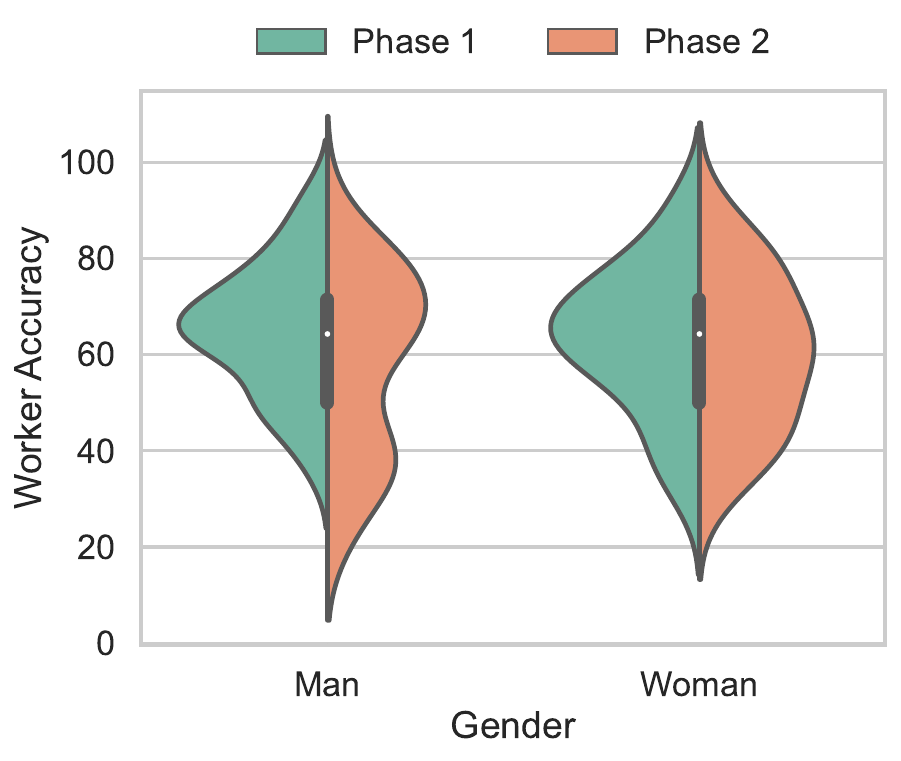}
         \caption{}
         \label{fig:gender}
     \end{subfigure}
     \begin{subfigure}[b]{0.48\columnwidth}
         \centering
         \includegraphics[width=.9\textwidth]{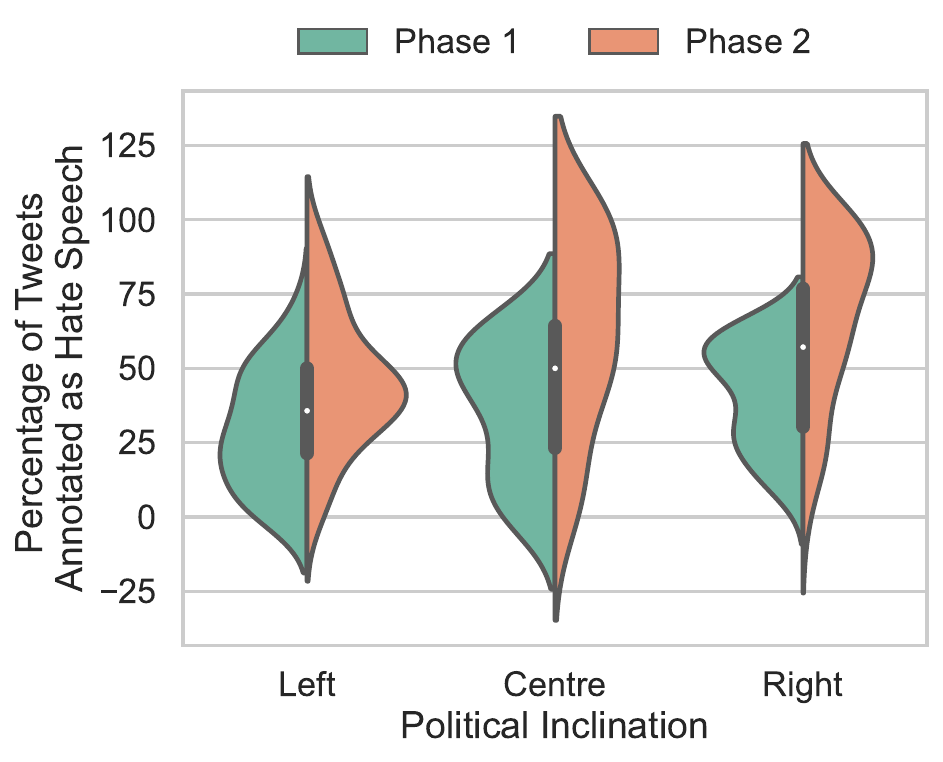}
         \caption{}
         \label{fig:political-hsp}
     \end{subfigure}
     \begin{subfigure}[b]{0.48\columnwidth}
         \centering
         \includegraphics[width=.9\textwidth]{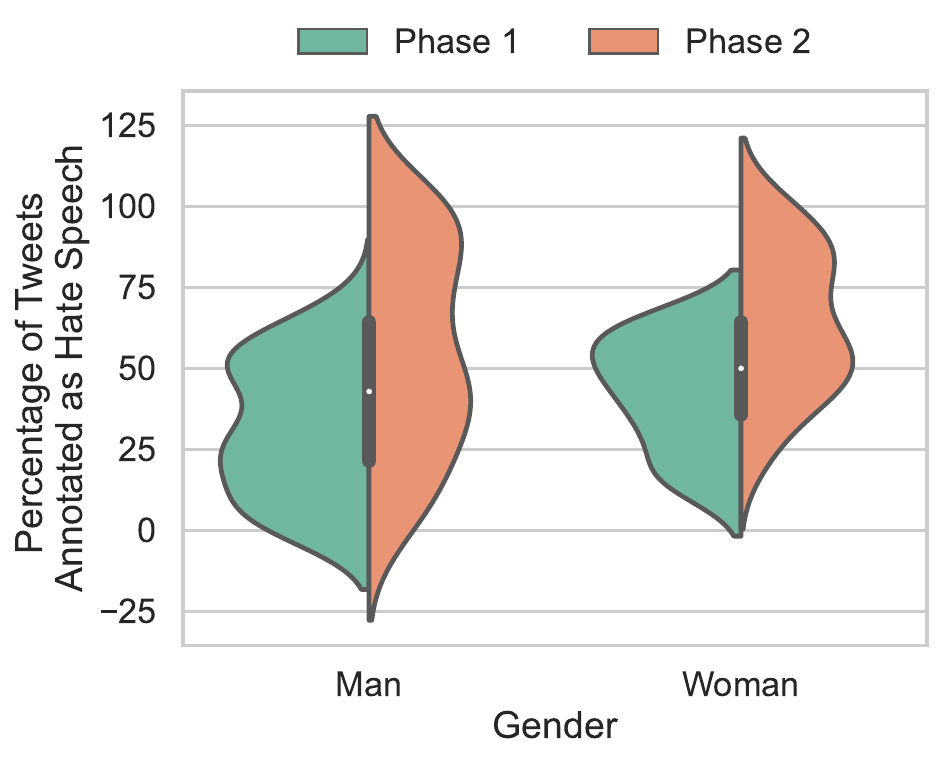}
         \caption{}
         \label{fig:gender-hsp}
     \end{subfigure}
     \caption{Impact of worker political inclination and gender on annotation outcomes.}
\end{figure}

Our results indicate a significant impact of worker political leaning on the hate speech annotation accuracy. As shown in {\bf Figure~\ref{fig:political}}, the difference is evident in annotations collected in both Phase 1 and Phase 2. In Phase 1, a Kruskal-Wallis test indicates that the worker accuracy differed over political leaning, $H(2) = 14.2$, $p < 0.001$. A post-hoc Dunn test with Bonferroni adjustment shows that worker accuracy for left-leaning workers ($N=39$, $M=70.3$, $SD=11.5$) is significantly higher compared to both centre ($N=23$, $M=57.5$, $SD=10.7$, $p<0.001$) and right ($N=47$, $M=62.6$, $SD=15.6$, $p<0.001$) leaning workers. There is no significant difference among the centre and right-leaning annotators. Similar results are evident in Phase 2, with the Kruskal-Wallis test showing a significant difference in accuracy based on the political leaning, $H(2) = 19.6$, $p < 0.001$. A post-hoc Dunn test with Bonferroni adjustment indicates a significantly higher accuracy for left-leaning annotators ($M = 68.7$, $SD=15.7$) compared to centre ($M=52.5$, $SD=15.6$, $p<0.001$) and right ($M=53.6$, $SD=17.7$, $p<0.001$) leaning workers, while the difference among the centre and right workers is not significant. 

\rev{In Phases 1 and 2, worker political leaning also significantly impacts the percentage of hate speech annotations provided ({\bf Figure~\ref{fig:political-hsp}}). In Phase 1, a Kruskal-Wallis test indicates that the percentage of hate speech annotations significantly varied over worker political leaning, $H(2) = 8.4$, $p < 0.05$. A post-hoc Dunn test with Bonferroni adjustment shows that the percentage of hate speech annotations provided by left-leaning workers ($M = 28.9$, $SD=19.7$) is significantly lower than right-leaning workers ($M = 41.5$, $SD=17.8$, $p<0.05$), and the differences among centre-right and centre-left are not significant. In Phase 2, a Kruskal-Wallis test indicates significant differences ($H(2) = 15.5$, $p < 0.001$), and a post-hoc Dunn test with Bonferroni adjustment shows that left-leaning workers ($M = 45.4$, $SD=22.5$) marked a lower percentage of tweets as hate speech compared to both centre ($M = 62.1$, $SD=32.6$) and right ($M = 68.1$, $SD=27.7$) leaning workers. The difference between centre and right workers is not significant.}

As seen in {\bf Figure~\ref{fig:gender}}, we further examine whether the self-reported gender of the worker impacts hate speech annotation accuracy. Mann Whitney U tests indicate no significant differences in accuracy based on the gender of the worker in both phases. In Phase 1, men ($N=65$, $M=65.1$, $SD=13.6$) demonstrated a slightly higher mean accuracy compared to women ($N=44$, $M=63.1$, $SD=15.2$), whereas in Phase 2, mean annotation accuracy for men ($M=57.9$, $SD=19.2$) was lower than for women ($M=60.0$, $SD=16.3$). 
\rev{In Phase 1, a Mann-Whitney U test indicates a significantly higher percentage of tweets tagged as hate speech by women ($N=44$, $M=42.4$, $SD=17.1$) compared to men ($N=65$, $M=31.4$, $SD=21.0$) ($U=995.5$, $p<0.01$). The difference is not statistically significant in Phase 2 ({\bf Figure~\ref{fig:gender-hsp}}).}

\subsection{Qualitative Results}

In this section, we draw on qualitative data collected through semi-structured interviews with nine \mturk\ workers. Participant demographics are shown in {\bf Table~\ref{tab:participants}}. In exploring the sensemaking processes of online crowd workers, we were especially inspired by \acite{Miceli2020-qt}'s investigation of the work of data annotation for computer vision, structured around three questions: ``How do data annotators make sense of data? What conditions, structures and standards shape that sense-making praxis? Who, and at what stages of the annotation process, decides which classifications best define each data point?'' (p. 2). The qualitative component of our research set out to address similar guiding questions: How do crowd workers make sense of hate speech? What conditions, structures and standards shape that sense-making? and How, and at what stages of the annotation process, do crowd workers decide what comprises hate speech?

\begin{table}[b]
\footnotesize
    \centering
    \begin{tabular}{lllll}
    \hline
        ID & Gender & Sex & Age (years) &  Crowd worker for  \\
        \hline
        P-1&	Man	&Male	&18-30	&More than 4 years\\
        P-2&	Man	&Male	&30-40	&More than 4 years\\
        P-3&	Man	&Male	&30-40	&1-2 years\\
        P-4&	Woman	&Female	&40-50	&Less than 1 year\\
        P-5&	Woman	&Female	&30-40	&2-3 years\\
        P-6&	Man	&Male	&60-70	&More than 4 years\\
        P-7&	Man	&Male	&18-30	&Less than 1 year\\
        P-8&	Woman	&Female	&40-50	&More than 4 years\\
        P-9&	Man	&Male	&18-30	&1-2 years\\
        \hline
    \end{tabular}
        \caption{Demographics of the participants.}
        \label{tab:participants}
\end{table}

Interviews took approximately 45 minutes and were conducted via remote video meetings. The first half of the interview covered questions regarding each worker's previous experience of crowd work in general, and annotation tasks in specific. In the second half, the worker was invited to `demonstrate how they [do] their work' ~\cite{Gray2016-fi} by sharing their screen and talking through the decisions that underpin their work practices. Participants were asked to describe how and why they make certain decisions to better understand the context underscoring their decision-making, and whether and how these contexts were individually or structurally determined. 

With participant consent, interviews were recorded, including both audio and visual feeds. Participants were invited to switch their camera off if they desired, but the interviewer kept their camera on throughout the interview. 
Following \acite{Braun2006-uy}, we undertook a thematic analysis, identifying descriptive themes or patterns within the qualitative data. Our analysis was inductive (driven by the data) rather than deductive (theory driven). As such, we used an open coding approach~\cite{Holton2007-ao}. 

Our analysis was multi-staged. First, one of the authors familiarised themselves with the data by reading and re-reading the transcripts. They then coded the data, looking for themes and patterns. Third, they collaborated with another author to review and refine the codes into larger, descriptive themes. In undertaking this process collaboratively, we sought not to seek replicability or objective accuracy, but rather internal consistency and representation of experiences within and across the corpus.

In what follows, we show how workers made sense of the task, and how this sensemaking was shaped by complex negotiations between subjective motivations, platform structures, task instructions, and external contextual factors. In doing so, we highlight the ways in which the outcomes of crowd working are influenced by the structures under which they labour. In making this argument, we move away from the existing focus on subjective influences of worker characteristics and their impact on biases stemming from annotator beliefs, characteristics, and demographics, and towards a more holistic accounting of influential factors.

\subsubsection{Defining hate speech}

In discussing how they defined hate speech, and thus categorised the tweets, participants frequently opined on the importance of free speech, drawing on geographically and culturally bound understandings of hate speech. According to P-3 (\ie~Participant 3), for instance, \qt{I think you should be able to say anything you want for the most part, as long as it's not threatening [… I'm] a pretty free speech type of person, to be honest}. Likewise, P-4 said, \qt{I'm more in favour of free speech. So even if it might be offensive to people, I think people should still be able to say it}. As P-4 continued: \qt{I feel like people can say stuff, in America at least}. P-4 stated \qt{[…] In the USA, we have different laws, I believe, for hate speeches}. In these instances, definitions of hate speech were geographically (American) and culturally (Western) bound. For P-9, these influences were evidence of the collective generation of such definitions: \qt{we're all influenced by each other, whether we like it or not}. The importance placed on free speech directly influenced how participants approached annotation tasks. As P-8 explained: \qt{I would prefer to train it [through annotations so]  there's like leeway, not everything is meant to be offensive, even if it accidentally is}. P-9 echoed this sentiment, noting that \qt{hate speech is too ambiguous. It's too all-encompassing and it's like a thinly veiled attempt at censorship, and it's clearly taking sides}. To deviate from this view, and to classify something as hate speech, then, P-3 told us that \qt{you have to [be able] to say there's definitely, the person hates this person}, or that there was clear evidence of \qt{hate speech towards, like, a race or a gender}, or what P-8 described as \qt{a protected class type vibe}. 

Other participants reiterated similar views and told us how previous experience with similar annotation tasks, and/or awareness of legal approaches to understanding hate speech informed their work. P-4, for example, told us that: \qt{I've kind of been conditioned by the past tasks to assume that true hate speech should have implicit or explicit calls to violence, or be a little bit more than just politically incorrect}. 

Fluctuations in hate speech sensemaking occurred when such `standard' definitions of hate speech encountered personal values or subjective ethos. P-1, for example, used empathy to interpret whether a tweet was hate speech or not: \qt{I'm of the belief that we don't need to have something happen to us to empathise with whether or not it's right or wrong}.  These values influenced annotating practices to varying degrees. For P-8, personal values took precedence over following task instructions or established definitions, sharing that \qt{Sometimes I get in trouble for it, but I think it's worth it. […] I mean there's stuff worth planting your feet in the ground for}. For some crowd workers, their subjectivity was an inherent part of their annotation work, with this something they presumed task requesters were taking into account. As P-6 told us, \qt{I bring my experience to the table, and I figure that's part of what you're paying for}. For this participant, tasks had to be congruent with their personal values, or else they would \qt{sign out of them and go on something else} altogether. 

At multiple stages throughout the experiment, participants encountered `borderline' cases that introduced significant ambiguity. In these instances, a range of different values and information-based strategies were employed. Those guided by values were most likely to err on the side of caution to mitigate potential instances of hate speech. As P-6 stated, \qt{My feeling about hate speech is if it makes somebody uncomfortable, why do it? And if you're doing it after somebody is uncomfortable, then it's on you. And so you err on the side of caution.} In this instance, the participant's values oriented their consideration of others, and led to their tendency to lean towards categorising tweets as hate speech, rather than not.

\subsubsection{Weighing up financial imperatives, task instructions, and subjective values} 
\label{sec:rewards-instructions-values}

The influence of financial imperatives on the sense-making practices of our participants cannot be overstated. For all nine of our participants, the use of \mturk\ was principally driven by financial incentives. P-3, for instance, had taken up crowd working when one of their existing three sources declined: 
\qt{when I first started it \mturk\, I had just wanted another source of income, because I had three sources, [and] one became less. So, I'm like, okay, I'll check it out […] it's a nice little source of a little extra [income], not the greatest, but it's okay}.
 
Despite the strong influence of personal values for some participants just described, in most instances, task instructions were prioritised. For example, while P-4 noted that while their \qt{subjective reasoning} underlay the sense-making that informed their annotations, these were superseded by the task instructions: \qt{the task instructions are always my primary goal}. The primacy of task instructions was driven by financial imperatives: \qt{At the end of the day, this is a job I'm getting paid in actual physical currency. So, I do try to do a good job as much as possible}. As P-4 explained: \qt{they [task requesters] are not looking for my opinion. They have a set of guidelines, and they want their data in those buckets. And I do take this job seriously […] I don't want to screw up someone's data set}. If participants deviated from the task instructions, they told us, they risked not receiving payment for their labour. This was reiterated more strongly by others, like P-9, who told us that \qt{being subjective is a big no-no for me on \mturk\ because there's a lot of people who don't pay you for your batches due to subjective criteria}. 

For some participants, task instructions were followed even if these were at odds with personal and subjective values: \qt{usually, they have rules on what they consider hate speech in the instructions. […] So you follow their rules. Do I think it's always hate speech? No, but uh, you follow the rules} (P-8). P-9 adopted a similar outlook in stating that in doing annotation tasks, \qt{I'm gonna do what my overlord's telling me to, and I'm gonna get my money and I'm gonna be on my way}. A notable exception was P-6, who articulated that due to their experience and accumulated status on the platform, they could adopt more of a subjective viewpoint, \qt{You've got a bit of power in the system to have your own stance, I suppose}. 

The details included in task instructions were, therefore, critical for participants to negotiate tensions between their subjective values and their need to get paid. For example, for P-1, if the task instructions said simply: \qt{`annotate hate speech', then I'm gonna be like, `okay, what do I think of hate speech as?' But if it says, `annotate hate speech, here's your definition, here is the lens you should be looking at it through', then I would obviously follow the instructions there.}. Broader understandings of the type of annotations expected by requesters also influenced how workers interpreted instructions, as identified by P-2, \qt{If I'm unsure, I revisit the criteria that's been laid out}.  For P-8, these kinds of tasks required additional focus. As they explained, when working on annotation tasks with specific guidelines, they would \qt{end up getting out my handy dandy notebook and making notes of what the instructions are}. This was because, \qt{if you don't follow instructions, you won't get any more work from that requester, or your task could be rejected} (P-8). 

Reflecting on the lack of definitive instructions in Phase 1, P-4 commented that they would usually \qt{feel uncomfortable} at this point and would \qt{message the [task] requester and ask them for more guidelines}. According to this participant, achieving this clarity was critical: \qt{the more communication, the more successful the project will be}. This was due to prior experiences, where task requesters had provided insufficient guidance, and the resulting annotations were rejected. For P-4, clarity around task instructions was critical to ensuring requesters received \qt{high-quality data}, and that the crowd worker received payment for their labour. As articulated by P-8, \qt{I mean, it's whatever you want, you're paying for it, so you should get what you're paying for.}

\subsubsection{Filling in the gaps}

All nine participants commented on the difficulty of the task at hand, specifically referring to the challenge of categorising something as hate speech with minimal context, such as the surrounding conversation, or information regarding the sender and recipient. This was especially so in instances of `borderline' cases. As P-3 explained, \qt{I don't know the context, so it's a little hard}. For P-8, decision-making in these instances was further challenged by what they described as the combination of their age and little social media presence. As a result, they were sometimes unsure if a word or phrase was offensive or not. To negotiate this, they described workaround practices, such as visiting websites like Urban Dictionary, where they would endeavour to decipher what was being said. As they explained, \qt{I'm not on social media a whole lot, I felt like, maybe these are terms that people are using […] maybe I'm out of the loop}. Participants also noted that their ability to effectively complete such tasks relied on their keeping up with evolving understandings of hate speech. As reflected by P-2, \qt{like ten years ago, that might not be classified as hate speech, but now that might be […] things have changed}.

The gap between the subjectivity of the individual crowd worker and the recipient of the hate speech also requires negotiation through the annotation process. While P-3 noted that although the text of the tweet in question \qt{doesn't really irritate me […] it feels like hate speech just because it feels really aggressive}. In this instance, P-3 told us, they would have preferred to categorise this instance within a third category to indicate its borderline status. Given, however, that this stage of the study allowed participants only to categorise tweets as `hate speech' or `not hate speech', P-3 ultimately decided \qt{I'd say hate speech}. In later stages of the study, the examples provided within the task instructions, were seen as useful for demystifying ambiguous cases. As P-1 explained, \qt{a set of examples is always helpful}. Moving beyond the influence of task instructions and the subjective experience of the crowd worker, P-8 also took into account societal discrimination they \qt{observe in the world} in determining whether something was hate speech or not.

To aid with this process of discernment, multiple participants described looking for key words or a \qt{handful of phrases or a handful of keywords that jump out at you} (P-1). In a similar vein, P-4 told us that they \qt{look for specific words}. Re-reading for context surrounding the perceived hate speech served a similar purpose for P-2: \qt{[…] this section right here jumped out for me right away, and I went and reread it again to understand the full context}. Referring to task instructions also provided participants with scaffolding for understanding and categorising tweets as hate speech or otherwise. P-4, for instance, revisited the task instructions to remind themselves of what \qt{specific racial slurs} were referred to. While the task instructions were useful for this participant in terms of filling in the gaps, for others like P-4, the additional detail introduced further challenges and increased the time required to complete the task: \qt{obviously, there's going to be more thinking that goes to it}.

\subsubsection{Accounting for contextual factors}

Beyond the influence of subjectivities and platformed structures such as financial incentives, participants also told us how their annotations could be influenced by external contextual factors, like the mood that they were in that day. As P-3 explained, 
\qt{Depending on my mood that day, I might be `I hate how sensitive the whole world is today', and I might be like, `okay, I'm [going to select] non-hate speech'. It could be my mood, something I read, [or] just how I feel about things that day that could affect the way I'm answering}.

Annotation tasks could have affective impacts on the crowd workers themselves. P-4, for instance, told us about hate speech annotation tasks that they had worked on previously. Unlike in other tasks, where they \qt{would just keep going}, with hate speech annotation, they had to \qt{walk away from my computer every 30 minutes or so. Because even if you can be emotionally disconnected, it just starts to fatigue you because it's exhausting}. Such emotional labor underlying both volunteer and commercial moderation work has been well-documented~\cite{Dosono2019-xh,Gilbert2020-nk,Roberts2019-xt,Steiger2021-pk,Wohn2019-te}.

\section{Discussion}

\subsubsection{Annotations and Worker Characteristics}

Our results indicate a significant relationship between crowd worker annotation accuracy and worker characteristics. As detailed in Table~\ref{tab:correlations}, we note a strong negative correlation between worker moral integrity, measured by the moral identity questionnaire and hate speech annotation accuracy. Moral identity indicates one's desire to make moral intentions and actions consistent, in other words, equality of private and public action. Guided by our qualitative findings in the sub-section on financial imperatives, we argue that following task instructions and providing accurate (\ie~expert-aligned) subjective judgements in hate speech annotation tasks require workers to suppress their values, leading to the observed negative correlation. The impact of moral identity also aligns with work by \acite{Hube2019-nv} where workers with strong opinions provided biased labels. 

In prior work, the personality of crowd workers is known to be associated with objective task performance~\cite{Lykourentzou2016-jf}. Similarly, our results indicate moderate positive correlations between Phase 2 accuracy and agreeableness, conscientiousness and openness. In line with our findings, \acite{Sang2022-an} reported that annotators' big five personality factors, including agreeableness and conscientiousness, influence how workers judge hate speech annotations of `showing fear', which include misogyny, homophobia, and xenophobia. \rev{Our results related to Ambivalent Sexism Inventory suggest that a higher score of benevolent sexism (\ie~valuing stereotypical attributes in women and believing that women need to be protected) could lead to tagging more content as misogynistic hate speech and subsequently reducing the overall task accuracy.}

Furthermore, our findings related to annotator gender agree with previous work finding no significant differences when men and women annotated hate speech directed at women~\cite{Wojatzki2018-lk,Goyal2022-hk}. \rev{However, women tend to mark more content as misogynistic hate speech than men in Phase 1, where additional categories, specific guidelines and examples were unavailable.}

Our findings are valuable to pre-select workers or assign tasks~\cite{Hettiachchi2020-bc,Barbosa2019-ln} to obtain expert-aligned outcomes when crowdsourcing hate speech annotations. However, we note that pre-selection based on demographics without evidence could lead to further biases. Thus, we recommend informed pre-selection and recruiting diverse crowds considering other factors. 

\subsubsection{Subjective Judgements}

Analysis of qualitative data from semi-structured interviews with crowd workers provides useful nuance to further understand the quantitative data just discussed. Annotation outcomes, and the process taken to achieve them are not only impacted by worker factors but simultaneously shaped by the structures under which they labour. In this way, our work further strengthens findings around crowd work that evidence the impact of structures and hierarchy on both crowd work labour and outcomes~\cite{Miceli2020-qt}.

The structures and standards, such as financial incentives and task instructions under which crowd workers undertake content moderation tasks shaped the process of defining hate speech, and the processes workers took to complete different tasks. For example, platform approval ratings, projected task earnings, fear of tasks getting rejected and understandings of what specific task requesters were looking for prompted workers to prioritise task instructions over personal values. While subjective values, empathy and altruistic motivations influenced some participants' labour~\cite{Miceli2020-qt}, the financial incentives for successfully aligning annotation outcomes with task requester expectations superseded individual values for the majority. 

Similar to objective tasks~\cite{Gadiraju2017-fy}, clarity in task instructions and examples appear to play a vital role in subjective judgements. Our quantitative results in Figure~\ref{fig:tweets-category} revealed that the percentage of tweets accurately tagged as hate speech in Phase 2 is higher compared to Phase 1. In our qualitative findings, we highlighted that workers expect detailed task instructions for subjective tasks. However, workers are also primarily driven by financial needs and want to maximise the rewards they receive for work duration. Therefore, task requesters need to be mindful of the additional time required to read and understand instructions. While more detailed instructions can cover more borderline cases, there is a potential risk of workers ignoring or skimming them if they are too time-consuming.

Surfacing the complexity of these negotiations helps to clarify the structural influences that impact the crowd working process and resulting hate speech annotations. In many of the cases just described, these structures overshadowed the impact of individual worker characteristics. Moreover, hate speech sensemaking was culturally and geographically bound, with legal definitions and broader personal expectations affecting workers' interpretations of hate speech \cite{Miceli2020-qt}. Although crowd work is often discussed in terms of its geographic dislocation ~\cite{Elbanna2022-zd}, our findings indicate that geography may remain relevant in terms of defining hate speech. In outlining these findings, we indicate the potential for a more holistic accounting of influential factors on crowd work annotations, and raise critical questions concerning the links between task requesters, platform structures, and worker practices, that underpin crowd work outcomes. 

\subsubsection{Limitations}

We note three key limitations in our work. First, avoiding potential biases and increasing the variety of hate speech in the dataset was an important consideration~\cite{Rahman2021-pt}. Thus, we avoided an approach like keyword search and used a set of seed profiles. However, with the current process, our hate speech dataset only consists of tweets relating to seven public figures. Second, we recruited only nine participants for the qualitative study due to challenges in recruiting crowd workers for interview tasks. Third, in a crowdsourcing setup, it is challenging to validate self-reported demographics and questionnaire outcomes~\cite{Ipeirotis2010-ak, Marshall2013-en, Ross2010-bz}. While we used several quality control measures, we acknowledge that there could be inconsistencies with collected worker attributes and questionnaire outcomes.

\section{Conclusion}

This work investigates how crowd workers annotate misogynistic and sexist hate speech. Through a \mturk\ study with 109 workers, we report a significant relationship between hate speech annotation outcomes and workers' political affiliation, moral integrity, benevolent sexist attitudes and personality traits. Furthermore, through interviews with nine workers, we provide detailed insights into how crowd workers make sense of hate speech. We also analyse conditions, structures, and standards that shape the sense-making process, and at what stages crowd workers make their labelling decisions during the annotation process. Overall, our findings inform design considerations for avoiding undesired biases in crowdsourcing subjective judgements, with the potential to enable more robust and fair automated moderation tools and thereby engender safer online spaces.

\section{Acknowledgments}

This research was conducted by the ARC Centre of Excellence for Automated Decision-Making and Society (CE200100005), and funded by the Australian Government through the Australian Research Council. We thank the participants for their valuable contributions.

\bibliography{ref}

\end{document}